\newcommand{\beq}{\begin{equation}}
\newcommand{\eeq}{\end{equation}}

\newcommand{\gc}{<0|(G^a_{\mu\nu})^2|0>}
\newcommand{\lqc}{\Lambda_{QCD}}

\documentclass{ws-procs9x6}
\begin{document}

\title{Confining fields in lattice $SU(2)$}

\author{V.I. Zakharov}

\address{Max-Planck Institut f\"ur Physik, \\
F\"ohringer Ring, 6 \\ 
80805 M\"unchen, Germany\\ 
E-mail: xxz@mppmu.mpg.de}


\maketitle

\abstracts{We review some difficulties of the standard picture
of confining fields which are viewed usually as `mild', or quasiclassical
 field configurations. The paradoxes are naturally resolved by
recent lattice observations of selftuned fluctuations which exhibit
both ultraviolet and infrared scales. The ultraviolet scale is
provided by the lattice spacing $a$ and is manifested in the non-Abelian 
action density associated with the fluctuations
while the physical scale is exhibited by their geometrical characteristics.
The data might suggest existence of a dual description of Yang-Mills theories
on the fundamental level. }

\section{Introduction}

Vacuum fields  are responsible for the confinement. This dogma is
rarely put in doubt. However, much less is known specifically on
the corresponding field configurations. Still the common belief is that
the confining fields are non-perturbative \footnote{See, however,
\cite{gribov}.} and `mild', see, e.g., a review 
\cite{dokshitzer} and references therein. The evidence is , however,
indirect. Moreover, in the Minkowski space it is difficult actually
to formulate what `soft' field means because the vacuum should look
the same in all the inertial frames.

The problem seems to be much more tractable in the Euclidean space.
One is inclined to think of a single size characterizing one or another
fluctuation. Moreover, there are
two intrinsic scales in QCD, ultraviolet and infrared. We will have
in mind the lattice regularization and by the ultraviolet cut off will
understand, therefore, the lattice spacing $a$. The infrared scale 
is provided by $\Lambda_{QCD}$.

There are well known examples which demonstrate relevance of these scales to 
the vacuum fluctuations. First, consider zero-point fluctuations. They have
no intrinsic scale built in and the typical size is
determined by an external probe. For example,  the vacuum
expectation value of the gluon condensate 
(that is, a point-like probe) is given by:
\beq\label{first}
<0|(G^a_{\mu\nu})^2|0>~\approx~{(N_c^2~-1)\over a^4}~~,
\end{equation}
where $G^a_{\mu\nu}$ is the gluonic field strength tensor and $N_c$
is the number of colors,
and is clearly ultraviolet-dominated. 
The very fact of the ultraviolet divergence in
the matrix element (\ref{first}) is not specific, of course,
for the lattice and well known from calculations of the vacuum energy
density in any formalism.
The lattice allows to unambiguously fix (\ref{first}).

On the other hand, 
the standard image for non-perturbative fluctuations is provided by
instantons. 
The typical size of the instantons is of order $(\Lambda_{QCD})^{-1}$.
Moreover
there exist detailed models of the instanton liquid developed along these lines
\cite{shuryak}.

Thus, the standard picture is that at short distances zero-point
fluctuations dominate while the confinement is due to soft
quasiclassical fields, something like instantons. This picture can be
probed, for example, through the sum-rule technique \cite{svz}
and its generalizations 
and claims many phenomenological successes, see, e.g., \cite{shuryak}
and references therein.

However, there has been accumulating evidence that this standard picture
misses sometimes even large, leading  effects
\cite{nsvz}. We will briefly review the issue in Section 3.
 In Section 4 we argue that the problems of the standard picture are
resolved if one takes into account the newly discovered {\it branes} 
\cite{kovalenko}. The branes are two-dimensional surfaces whose
total area scales in the physical units while the action density
is of perturbative order (\ref{first}). The branes represent a new 
kind of  {\it selftuned} fluctuations which unify the scales
$a$ and  $\Lambda_{QCD}$. 
This, written version 
of the lectures presents only an overview of the actual content
of the lectures given at the Summer Institute. 
Further details and references can be found in
other talks of the author \cite{vz}.

\section{Difficulties of the standard picture}

{\it Confinement} 

We are considering pure Yang-Mills theory and the criterion of confinement 
is the existence of a linear potential at large distances for heavy 
external quarks. The lattice data can indeed be fitted by a very simple
form,
\beq\label{cornell}
V_{\bar{Q}Q}(R)~\approx ~ -{const\over R}~+~\sigma\cdot R~~,
\end{equation}
at all the distances.  

The point crucial for our discussion here is that the instanton 
liquid model does not reproduce the linear piece at all \cite{confinement}.
Thus, the successes of the model in other cases turn into a problem for the
standard picture. Indeed, the model seems to be the best realization
of the standard picture but still misses the confinement.

{\it Current correlators}

A standard way of probing vacuum fields is 
provided by measuring current
correlators, $<0|j(x),j(0)|0>$ where $j(x)$ are local currents constructed
from the quark and gluon fields, see, e.g., \cite{svz,shuryak}.
The strategy is to start from short distances, $x\ll \Lambda_{QCD}^{-1}$
and approach `intermediate' distances where non-perturbative effects 
become sizable. In the most simplified form the theoretical
predictions look as:
\beq\label{simplified}
<0|j(x),j(0)|0>~\approx~({\rm parton~model)}\big(1~+
c_jx^4\cdot\gc_{soft} ~+...\big)~~,
\end{equation}
where the coefficient $c_j$ depends on the current considered and calculable
while $\gc_{soft}$ stands for the contribution of the soft non-perturbative
fields, $\gc_{soft}\sim\Lambda^4_{QCD}$. 
In actual applications, there can be other  vacuum condensates
representing the non-perturbative fields. Another variation
of (\ref{simplified}) is an explicit calculation of a single-instanton
contribution \cite{shuryak}. Also, one commonly keeps
the first-order perturbative contribution of order $\alpha_s(x)$
as well.

In many cases, the approximations like (\ref{simplified}) work well.
However, there are cases when (\ref{simplified}) fails,
see in particular \cite{nsvz}. Moreover, there is accumulating
evidence that in cases when the Born (parton-model) approximation is 
dominated by gluon propagator the leading correction is of order
\beq\label{quadratic}
 <0|j(x),j(0)|0>~\approx~({\rm parton~model)}\big(1~+
b_j\Lambda_{QCD}^2\cdot x^2~+...\big)~~,
\end{equation}
for the latest example of this kind and references see \cite{narison}.

There is no model-independent way to 
relate the  coefficients $b_j$ in (\ref{quadratic}) in various channels since
the quadratic correction  is not captured by the
operator product expansion. Phenomenologically, however, 
the model with a non-vanishing short-distance gluon mass $m_g$ 
(see second paper in Ref \cite{nsvz}) turns to be
successful. Within this model one modifies the propagator
of a gluon by replacing 
\beq\label{mass}1/q^2~\rightarrow~1/q^2 ~+~|m^2_g|/q^4~~,
\end{equation}
where $q$ is the momentum of the gluon, $q^2\gg m^2_g$. It is worth
emphasizing that the model (\ref{mass}) applies in the $Born~approximation$
when higher order corrections in $\alpha_s$ are ignored. (Modification
of the Born approximation valid at all $q^2$ has been introduced in 
\cite{thooft}). In particular, the
Cornell potential (\ref{cornell}) at {\it short distances} exhibits
a correction induced by (\ref{mass}). Indeed, the Born approximation
in this case is given by one-gluon exchange and this is just the case 
when one expects the dominance of the quadratic correction.

\section{Long perturbative series}

\subsection{Expectations}

The difference between `long' and `truncated' perturbative series
is crucial in view of the theorem \cite{beneke} that
terms of order $x^2\cdot\Lambda^2_{QCD}$ are calculable perturbatively.
Let us explain this point in more detail.

Generic perturbative expansion for a matrix element of a local operator
 looks as:
\beq \label{pert}
\langle~ O~\rangle~= ~(parton~model)\cdot
\big(1~+~\sum_{n=1}^{\infty}a_n\alpha_s^n~\big)~~,
\end{equation}
where we normalized
the anomalous dimension of the operator $O$ to zero.
Note also that $\alpha_s$ is the bare coupling,
$\alpha_s\ll 1$.

In fact, expansions (\ref{pert}) are only formal
since the coefficients $a_n$ grow factorially at large $n$:
\beq\label{growth}
|a_n|~\sim~c_i^n\cdot n!~~,
\end{equation}
where $c_i$ are constants. Moreover, there are a few sources
of the growth (\ref{growth}) and, respectively, $c_i$
can take on various values, for review see, e.g.,  \cite{review}.
The factorial growth of $a_n$ implies that the expansion (\ref{pert})
is asymptotic at best. Which means, in turn, 
that (\ref{pert}) cannot approximate
a physical quantity to accuracy better than
\beq\label{uncert}
\Delta~\sim~\exp\big(-1/c_i\alpha_s\big)~\sim~
\Big({\lqc^2\cdot a^2}\Big)^{b_0/c_i}~~,
\end{equation}
where $b_0$ is the first coefficient in the $\beta$-function.
To compensate for these intrinsic uncertainties one modifies
the original expansion (\ref{pert}) by adding the corresponding
power corrections
with unknown coefficients.

In case of the gluon condensate the theoretical expectations
can be summarized as:
\beq\label{expectations}
\langle 0|~{-\beta(\alpha_s)\over\alpha_s b_0}
\big(G_{\mu\nu}^a\big)^2|0\rangle~
\approx~\alpha_s{(N_c^2-1)\over a^4}\big(1+
\sum_{n=1}^{\sim N_{ir}}a_n\alpha_s^n~+~(const)a^4\cdot\Lambda_{QCD}^4\big)~~,
\end{equation}
where
$$N_{ir}~\approx~{2\over b_0\alpha_s}~~$$ 
and terms proportional to $\Lambda_{QCD}^4$ correspond to  
$<0|(G_{\mu\nu}^a)^2|0>_{soft}$, see (\ref{simplified}).
It is the latter quantity which enters the QCD sum rules, see \cite{svz}.
 
A conspicuous feature of the prediction (\ref{expectations})
is the absence of a quadratic correction,
compare (\ref{quadratic}).
However, Eqs (\ref{quadratic}) and (\ref{expectations}) are not necessarily
in contradiction with each other.
The point is that the phenomenological analysis 
relies on {\it short} or even very short truncated
perturbative series plus a power-like correction
while the theorem (\ref{expectations}) keeps 
a {\it long} perturbative series
(plus the first power-like correction)

\subsection{Numerical results}

Numerically, this problem was studied in greatest detail  in case of
the gluon condensate on the lattice \cite{rakow}. 
In terms of the lattice formulation the gluon condensate is nothing
else but the average plaquette action.  
The result can be summarized in the following way.
Represent the plaquette action $\langle P\rangle $ as:
\beq\label{plaquette}
\langle P\rangle ~\approx~P_{pert}^N~+~b_Na^2\lqc^2~+c_Na^4\lqc^4~~,
\end{equation}
where the average plaquette action $\langle P\rangle $
is measurable directly on the lattice and is known to high accuracy,
$P_{pert}^N$ is the perturbative contribution calculated 
up to order N:
\beq\label{pn}
P_{pert}^N~\equiv~1~-~\sum_{n=1}^{n=N}p_ng^{2n}~~,\end{equation}
and, finally coefficients $b_N,c_N$ are fitting parameters
whose value depends on the number of loops $N$. Moreover, the form of
the fitting function (\ref{plaquette})
is rather suggested by the data 
than imposed because of theoretical considerations.

The conclusion is that up to ten loops, $N=10$ it is the quadratic correction
which is seen on the plots while $c^N$ are consistent with zero.
However, the value of $b^N$ decreases monotonically with growing $N$ 
\cite{rakow}. The factorial divergence (\ref{growth})
is not seen yet and perturbative series reproduces
the measured plaquette action at the level of $10^{-3}$. 
Finally, at the level $10^{-4}$  the $\lqc^4$ term seems to 
emerge \cite{rakow}.

\section{Branes}

\subsection{Coexistence of two scales}

We have already  mentioned that instantons do not confine.
A natural question is then, what are the confining configurations. 
Actually the answer is also more or less known.
Namely it is commonly believed (for a recent review and references see
\cite{greensite}) that it is the monopoles and central vortices
that confine. The next question is then, why we did not account
for these fluctuations, say in (\ref{expectations}) or (\ref{simplified}).
The answer is pure technical, at first sight. Namely, the monopoles and 
vortices are defined in terms of projected, not original non-Abelian 
fields. The projection, in turn, is determined non-locally
and the track to the original Yang-Mills fields is lost.
However, the general expectation seems to be that the monopoles and
vortices correspond to soft, bulky fields and as far as dependence on
$\Lambda_{QCD}$ is concerned are not much distinguishable from the
instantons. 
 
These expectations turned to be not true.
Namely both the monopoles, see \cite{anatomy} and
references therein, and
vortices \cite{kovalenko} appear to be  associated with an excess of the
{\bf non-Abelian} action which is
divergent in the ultraviolet:
\beq\label{uvd}
\langle S_{mon}\rangle~\sim~\ln 7\cdot{L\over a}~,~~~
\langle S_{vort}\rangle~\approx~0.54\cdot{A\over a^2}~~,
\end{equation} 
where $L$ is the length of the monopole trajectory, $A$ is the
area of the vortex. 

It is most remarkable that the infrared scale is also relevant
to the branes. Namely, it has been known since some time (for review
see \cite{greensite}) that the densities of monopoles and vortices
scale in physical units. The corresponding densities are defined
as:
\beq\label{definition}
L_{perc}~\equiv~4\rho_{perc}V_4~~,~~A_{vort}~\equiv~6\rho_{vort}V_4~~,
\end{equation}
where $V_4$ is the volume of the lattice and $L_{pers}$ is the total length
of the percolating cluster while $A_{vort}$ is the area of the vortices.
The densities (\ref{definition}) scale in physical units and are
independent on the lattice spacing. According to the latest 
measurements:
\beq\label{ird}
\rho_{perc}~=~7.70(8)~fm^{-3}~~,~~A_{vort}~\approx~4.0(2)~fm^{-2}~~,
\end{equation}
see \cite{boyko} and \cite{kovalenko}, respectively. 
Moreover, the monopole trajectories lie on the P-vortices 
\cite{giedt,kovalenko}. 

\subsection{Selftuning}

Naively, one would expect that monopoles and vortices with action
 (\ref{uvd})
propagate only very short distances, $L\sim a$, $A\sim a^2$. However, 
both monopoles and vortices form
clusters which percolate through
the whole of the lattice volume $V_4$.
This implies cancellation of the ultraviolet divergences between action
and entropy. This cancellation is easy to quantify in case
of monopoles (particles). Namely, the propagating mass, $m$  of the monopole
is in fact {\it not} the radiative mass $M(a)$,where $M(a)\equiv S_{mon}/L$,
see (\ref{uvd}). The relation between the two masses is as
follows:
\beq\label{higgs}
m^2_{mon}~=~{const\over a}\big(M(a)~-~{\ln 7\over a}\big)~~.
\end{equation}
The $\ln 7$ term here is due to the entropy (see, e.g., \cite{ambjorn}).
Observation (\ref{ird}) implies cancellation between the action
and entropy.  Moreover, there is no parameter to tune to ensure this
cancellation. Thus, the monopoles
are rather {\it selftuned}. The same is true for the vortices.
Both the tension and entropy are ultraviolet divergent \cite{kovalenko}.

\subsection{Quadratic correction to the gluon condensate}

Monopoles and vortices are defined, for a given configuration of the
vacuum fields, for the whole of the lattice. Thus, they are seen 
as a nonlocal structure. Moreover, they are manifestly
non-perturbative. Indeed, 
the  probability $\theta(plaq)$ for a particular 
plaquette to belong to a brane
has been found to be proportional to:
\beq\label{area1}
\theta(plaq)~\approx~(const)\exp(~-1/b_0g^2(a))~\sim ~(a\cdot \lqc)^2~.
\end{equation}
On the other hand, the branes have an  ultraviolet divergent tension
which assumes a kind of locality.

 To make contact with the continuum theory it is useful to evaluate
contribution of the branes into local or quasi-local matrix elements.
The gluon condensate turns to be the easiest case. Indeed,
combining Eqs (\ref{ird}) and (\ref{uvd}) one gets for the
contribution of the vortices to the gluon condensate:
\beq\langle~(G_{\mu\nu}^a)^2~\rangle_{vort}~\approx~0.3~GeV^2~a^{-2}~~,
\end{equation}
We see that the mysterious $a^2\cdot \Lambda_{QCD}^2$ correction,
see Sect. 3.2.,
gets its explanation in terms of the branes. The mixture of the
the two scales, $a$  and $\Lambda_{QCD}$, exhibited by this correction 
appears to be a manifestation of selftuning of the branes.

\section{Status of theory}

\subsection{Why branes?}

The lattice data on the monopoles and vortices have been accumulating
since long. However,they were analyzed almost exclusively within the
confinement problem.
Discovery of the ultraviolet divergences, 
see (\ref{uvd}), makes the
challenge to the theory much more direct. Indeed  
in an asymptotically free theory one should be able to understand 
short distances from first principles. At present, however, 
the phenomenology is by far ahead of the theory.

Observation of the branes is gratifying
from the theoretical point of view. 
Indeed, the results (\ref{uvd}) imply that at least at presently available
lattices the size of the monopoles is not resolved. 
Moreover, gluons are already (approximately) free particles at such distances.
On the other hand, it is well known that
there exists no consistent field
theory with both electric (color) and magnetic dynamical charges. 
 Existence of the branes 
implies that the magnetic and electric (color) charges are separated
in space. The color charges live in the bulk while the magnetic charges
live on the two-dimensional branes (which percolate through the four-
dimensional Euclidean space).

\subsection{Constraints from asymptotic freedom} 

The asymptotic freedom implies that at short distances the only degrees
of freedom relevant are those of free gluons. How do we count the degrees
of freedom? Usually through ultraviolet divergences. The best known example
is the $\beta$-function which counts logarithmic divergences. The logarithmic
divergences are singled out since they are independent on details of the
cut off. However, on the lattice the power divergences are also 
uniquely defined. And they are much easier to study. In particular, if
we think in terms a scalar complex field $\phi_M$ describing monopoles
then we can introduce a vacuum expectation value $\langle |\phi_M|^2\rangle$
\footnote{The condensate can be thought of as non-perturbative part
of the gauge invariant condensate of dimention two, see \cite{stodolsky}.}.
For an elementary monopoles we would have 
$\langle |\phi_M|^2\rangle\sim a^{-2}$. This is not allowed since we 
may not have new particles at short distances. What is allowed is
\beq
\langle |\phi_M|^2\rangle
~\sim~\Lambda_{QCD}^2~,
\end{equation}
and this is a constraint from the asymptotic freedom. 

Remarkably enough, this constraint can be rewritten 
\cite{vz} in terms of the
{\it total} length of the monopole trajectories (in (\ref{definition})
we considered only the percolating cluster while now are 
including finite clusters as well \footnote{For further applications
of the percolation theory see , in particular, \cite{maxim,boyko}.}): 
\beq\label{surface}
\rho_{mon}^{tot}~\le ~(const){\Lambda_{QCD}^2\over a}~~.
\end{equation}
This constraint turns to be satisfied by the data \cite{boyko}.
On the other hand, from pure geometrical point of view
(\ref{surface}) implies that the monopoles actually percolate
on a two-dimensional surface, not on the whole of the four-dimensional
space. Which means branes.

In other words the data on the monopole action, see (\ref{uvd})
plus asymptotic freedom of the original YM theory imply existence of
branes. 

\subsection{Manifestations of the duality}

As we mentioned above, the original formulation of the YM theory 
does not explain the selftuning of the monopole action, compare
(\ref{uvd}) and (\ref{higgs}). Indeed the geometrical factor $\ln 7$
depends on the type of the lattice (cubic) and is meaningless in
the continuum. Imagine, however, that there were a theorem proving
existence of the dual formulation
in the continuum limit. Then the selftuning would be  derived since
it is a necessary condition for the monopoles 
to survive in the continuum limit.

Also, compare the contribution to the gluon condensate from instantons
and monopoles. The instantons are added to the perturbative expansion,
see (\ref{expectations}). The contribution of the branes, to the contrary,
cannot be added to the perturbation theory but is, instead, dual to the
perturbative series, see Sects. 3.2, 4.3. The reason is that the branes
`belong' to the dual world.

\section*{Acknowledgments}

I am thankful to the organizers of the Summer Institute, and 
especially to Prof J. Kubo and Prof H. Terao, for
the invitation and hospitality.
The paper was worked out while the author was visiting the Yukawa
Institute for Theoretical Physics of the Kyoto University.
I am thankful to Prof. T. Kugo and other members of the group for
the hospitality.



\end{document}